\documentclass[journal=apchd5,manuscript=article]{achemso}

\usepackage{chemformula} 
\usepackage[T1]{fontenc} 

\author{Marina Radulaski}
\affiliation{E. L. Ginzton Laboratory, Stanford University, Stanford, CA 94305, USA}
\alsoaffiliation{Hewlett Packard Enterprise Labs, Palo Alto, CA 94304, USA}
\email{marina.radulaski@stanford.edu}
\author{Ranojoy Bose}
\affiliation{Hewlett Packard Enterprise Labs, Palo Alto, CA 94304, USA}
\email{ranojoy_bose@apple.com}
\author{Tho Tran}
\affiliation{Hewlett Packard Enterprise Labs, Palo Alto, CA 94304, USA}
\author{Thomas Van Vaerenbergh}
\affiliation{Hewlett Packard Enterprise Labs, Palo Alto, CA 94304, USA}
\author{David Kielpinski}
\affiliation{Hewlett Packard Enterprise Labs, Palo Alto, CA 94304, USA}
\author{Raymond G. Beausoleil}
\affiliation{Hewlett Packard Enterprise Labs, Palo Alto, CA 94304, USA}

\title{Thermally tunable hybrid photonic architecture for nonlinear optical circuits}


\begin{document}

\begin{tocentry}
\begin{center}
\includegraphics{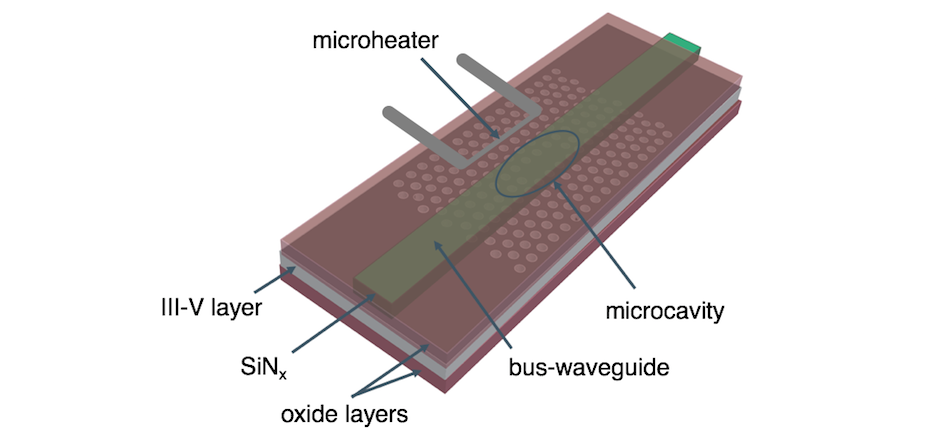}
\end{center}
\end{tocentry}

\begin{abstract}

Integrated photonic systems are a leading solution for fast and energy-efficient information processing. We develop a thermally tunable hybrid photonic platform for implementation of coherent optical networks applicable to dedicated computing and machine learning. The hybrid architecture comprises gallium arsenide (GaAs) photonic crystal cavities, silicon nitride (SiN$_x$) grating couplers and waveguides, and chromium (Cr) microheaters on an integrated photonic chip. We utilize the band-edge nonlinearity in GaAs photonic crystal cavity for switching at 6 ps timescale. The cavities are evanescently connected to a common bus waveguide, separating the computation and communication layers. To synchronize the operating frequency, we design metal microheaters that locally, continuously and reversibly tune photonic crystal cavities to a common resonance. We demonstrate 7 nm tunability range with no significant quality factor deterioration. By coupling the free-space light to the photonic chip, we demonstrate the tunable interaction of the propagating signal with three nonlinear optical nodes.

\end{abstract}

\section*{Introduction}

All-optical systems are a leading solution for low-power and high bandwidth digital and analog information processing with a broad range of applications in reservoir computing \cite{vandoorne2014experimental}, machine learning \cite{shen2017deep}, Ising networks \cite{wang2013coherent}, and special-purpose computing \cite{wu2014computing}. These optical systems bring together highly nonlinear nodes as fast switching and memory elements \cite{kuramochi2014large}, optical interfaces such as grating couplers and waveguides that allow connection between arbitrarily separated nodes, and tuning elements that provide complete control over the phase and the frequency of light propagating in the system. Many of these systems rely on optical phase relationships between the many different components in the optical circuit, which are enabled through the coherent control of light that propagates on chip. While silicon and amorphous silicon photonics benefit from CMOS compatibility and mature fabrication technology and are therefore attractive for building scalable optical circuits, they also suffer from drawbacks stemming from poor nonlinear optical performance. Devices made in bulk III-V materials on the other hand, have been shown to offer higher optical nonlinearities and operating speeds, taking advantage of enhanced free-carrier dispersion and linear absorption near the material band edge \cite{Bennett, nozaki2010sfa}. Photonic crystal cavities that confine light in very small volumes and remarkably enhance nonlinear optical effects are optimal as these nonlinear building blocks. However, connecting several cavities on a single optical chip via photonic crystal waveguides with control over the optical phase is challenging because of design and lithographic constraints associated with the highly ordered nature of photonic crystals. As an alternative bus waveguides that operate in a vertically spaced optical plane and evanescently couple to distant photonic crystals, can be used \cite{halioua2011hybrid, debnath2012cascaded, debnath2013highly, katsumi2018transfer}. These waveguides offer much higher flexibility in designing optical circuits. Device tunability can be enabled by metallic microheaters placed adjacent to the bus waveguide. Such a platform can be used to tune the frequency and relative phases of distant photonic elements arbitrarily, and is therefore a powerful tool for controlling the coupling between cavities on chip that are only achieved with difficult engineering design within traditional planar photonic crystal architectures \cite{Haddadi:14}.

The hybrid integration of III-V substrates with silicon or silicon nitride has been widely explored for applications such as quantum well lasers \cite{Bazin:14}, as well as for applications in quantum information processing \cite{Davanco2017}. These systems combine highly desirable optical properties of the III-V materials, for example, high internal quantum efficiencies of quantum wells and quantum dots, with desirable properties of the silicon-based materials, for example, low-loss optical propagation, Kerr nonlinearities, dispersion engineering, and a multitude of other linear and nonlinear effects.

Here, we present a fully scalable, thermally tunable, hybrid III-V/SiN$_x$ photonic platform for all-optical computing applications. Individual gates in our system incorporate highly nonlinear gallium arsenide photonic crystal cavities that are designed to enhance the nonlinear free-carrier effect through near-band-edge resonances for fJ-scale switching near 900 nm wavelengths \cite{bose2015carrier}. The photonic crystal cavities are clad using oxides below, above and inside the etched holes leading to stable devices and enhanced thermal management compared to air-bridged cavities, but continue to demonstrate optical switching at low energy and high speeds. Light in our optical circuit propagates in a layer vertically above this oxide layer via low propagation-loss SiN$_x$ bus waveguides, connecting several nodes that can be spatially separated by multiple wavelengths of light. We also achieve resonance tuning for the photonic crystal modes using chromium Ohmic microheaters designed for local and reversible resonance tuning of more than 5 nm, and integrated on chip. Using this platform, we demonstrate that multiple GaAs optical cavities can be probed using low-index SiN$_x$ waveguides, and modes of one cavity can be tuned through the resonances of other cavities. The developed architecture is widely usable in a range of optical circuit applications, and supports operation at high speeds and low energy. All experiments are performed at room temperature using bulk effects.

\section*{Hybrid photonic architecture}

The purpose of the developed hybrid photonic architecture is to support connections between multiple GaAs photonic crystal cavity gates that can serve as the basic nonlinear switch or memory element in arbitrary optical circuits. The nodes are buried in oxide layers providing sufficient contrast for high quality factor resonances. A SiN$_x$ bus-waveguide overlays the cavities offset by an intermediate layer of SiO$_x$ whose thickness of 200 nm has been selected to facilitate optimal coupling. This dimension is in agreement with literature \cite{katsumi2018transfer} where waveguide coupling strength is shown to significantly increase for oxide thickness of 200 nm and above. To tune the cavities into a common resonance, Ohmic microheaters are defined in a Cr layer on top of the glass.

\subsection*{Microheater and waveguide design}
The thermal tuning element plays a significant role in determining the performance of optical circuits. In addition to phase and frequency control, a microheater can help mediate fabrication errors in electron-beam lithography introduced by nano-scaled variations in devices that affect the uniformity of their performance. In the case of GaAs photonic crystal devices at 900 nm where typical hole dimensions are 120-140 nm, the resonances of identically designed photonic crystal (PhC) cavities in our platform can shift by up to 5 nm as a result of fabrication errors.

The frequency tuning of photonic crystal cavity resonances has been addressed though a variety of global or irreversible techniques. These include hot-plate tuning \cite{wild2004temperature}, chemical digital etching \cite{hennessy2005tuning}, laser-induced photo-oxidation \cite{piggott2014photo} and the chalcogenide glass method \cite{faraon2008local}. However, to compensate for the individual device resonance drift induced by fabrication, temperature fluctuation, or oxidation, local and reversible methods need to be implemented. Promising steps in this direction have been achieved through laser \cite{yang2009all, pan2010tuning} and microheater \cite{chong2004tuning, gu2007thermooptically, cui2012thermo} thermal tuning. The microheater approach, in particular, has an advantage in hybrid photonics integration because of electrical control and flexible positioning relative to the cavity region and other optical elements in its vicinity.

\begin{figure}[htbp]
\centering
\fbox{\includegraphics[width=\linewidth]{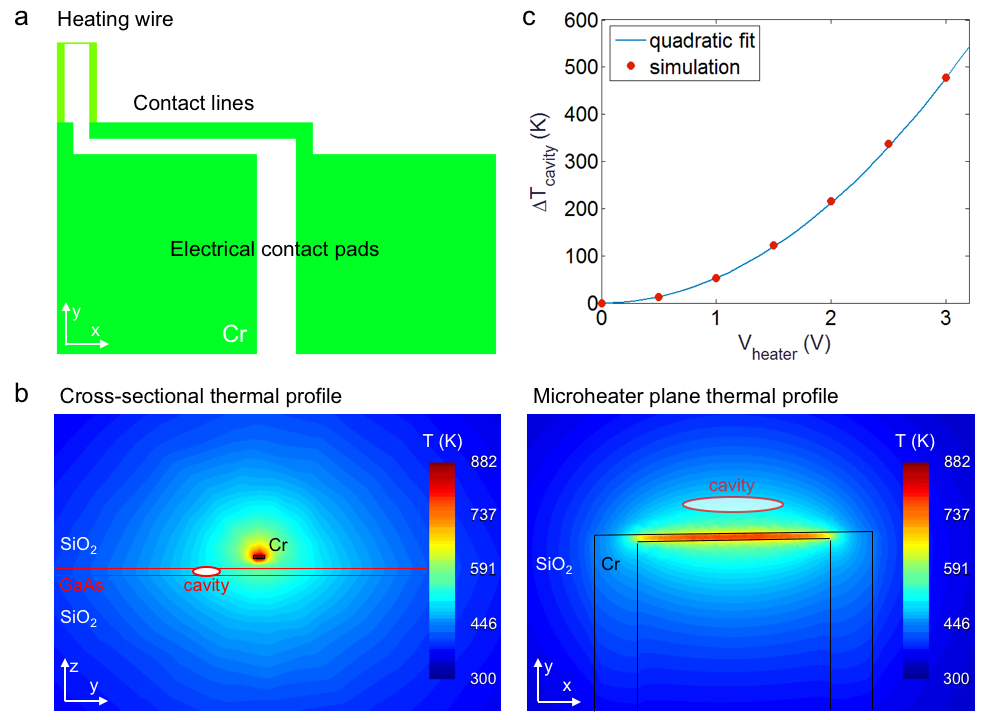}}
\caption{a) Microheater design comprises 50 $\mu$m x 50 $\mu$m electrical contact pads 1 $\mu$m wide contact lines and 200 nm wide k$\Omega$-range Ohmic heating wire. b) Modeled temperature profile shows that the heat is dominantly released in the Ohmic heating wire under 2 V DC drive; the red ellipses indicate the position of the PhC cavity in the buried GaAs layer. c) Projected temperature change in the PhC GaAs cavity as a function of the applied voltage.}
\label{fgr:heat}
\end{figure}

We design an Ohmic microheater that can generate hundreds of degrees of temperature change in a GaAs PhC cavity region, thus causing an index of refraction change of up to $\Delta n = 0.025$, and shifting the resonance by up to 5 nm. The microheater contains  three sections: the heating wire, the contact lines and the electrical contact pads (Fig. \ref{fgr:heat}a). The contact pads are 50 $\mu$m x 50 $\mu$m square areas designed to interface the electrical DC probes. The contact probes deliver the current to the 200 nm wide heating wire, which is laterally and vertically offset from the PhC cavity region. The heat transport simulation (Lumerical HEAT software) shown in Figure \ref{fgr:heat}b reveals the temperature profile of an electrically driven Cr heater in a SiO$_2$ substrate with a 200 nm thick GaAs slab placed 200 nm beneath the heater. To assure numerical convergence, we replaced air and the layer below the GaAs slab with SiO$_2$. We believe this alteration does not significantly influence the conclusions because the model captures the heat transport from the Cr nanowire to the GaAs slab through the SiO$_2$ layer, which is the main transport mechanism in the realized system. The simulated profiles show that most of the heat is dissipated in the heating nanowire whose resistance is in the k$\Omega$ range. About 30\% of the wire's temperature change is delivered to the buried GaAs slab at 1 $\mu$m lateral distance. The projected temperature tuning of the PhC cavity with the applied voltage is shown in Figure \ref{fgr:heat}c. The temperature change of 200 K corresponds to the refractive index change $\Delta n \sim 0.01$ \cite{mccaulley1994temperature}. The temperature gradient spans an 8 $\mu$m radius around the cavity providing a mechanism for localized reversible resonance tuning.

\begin{figure}[htbp]
\centering
\fbox{\includegraphics[width=\linewidth]{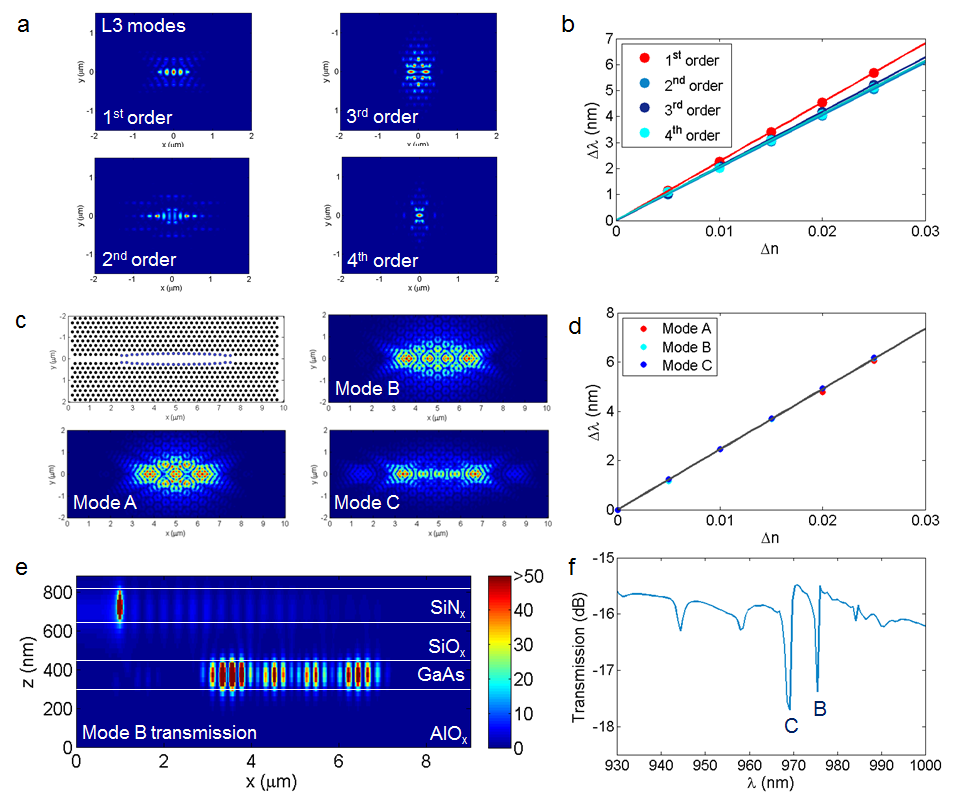}}
\caption{a) The FDTD electric field $|E|^2$ profile of four lowest energy modes of an L3 cavity. b) The modeled L3 resonant wavelength dependence on the index of refraction change showing a similar linear tuning trend for all modes. c) The design of the perturbed photonic crystal cavity with shifted holes indicated in blue and the FDTD electric field $|E|^2$ profile of three higher order modes, labeled Mode A, B and C, of the perturbed photonic crystal cavity. c) The modeled resonant wavelength dependence on the index of refraction change showing the same linear tuning trend for all modes. e) The FDTD $|E|^2$ profile of the light from the waveguide being coupled to the Mode B of the buried PhC cavity. f) The transmission spectrum of modal light transmitted through the waveguide and coupled to PhC cavity, with indicated Modes B and C.}
\label{fgr:tuning}
\end{figure}

Next, we theoretically analyze the GaAs photonic crystal cavity resonance response to the refractive index tuning expected from the integrated microheater. For this purpose, an L3 photonic crystal cavity \cite{akahane2003high} with lattice constant $a = 240$ nm and hole radius $a = 55$ nm buried in glass, and a perturbed PhC cavity with lattice constant $a = 240$ nm and hole radius of $r = 70$ nm are simulated using the Finite-Difference Time-Domain method (Lumerical FDTD software). The perturbed cavity is formed by symmetrically shifting the y-position of 44 holes in the $\Gamma-K$ direction of a photonic crystal waveguide \cite{debnath2013highly} (Fig. \ref{fgr:tuning}c). The y-shifts starting with the central four holes are [40, 35, 35, 28, 28,18, 18, -3, -3, -40, -40] nm away from the waveguide area. The cavity is surrounded with AlO$_x$ layers beneath, and SiO$_x$ layer on top and filling the holes, with refractive indices are $n_{AlO_x}=1.6$ and $n_{SiO_x}=1.4$. The spectral response of the systems is recorded as we modulate the index value of the entire cavity in increments of $\Delta n = 0.005$. The four lowest energy resonances in the L3 cavity are observed to tune linearly by 5-6 nm for $\Delta n = 0.025$ change in refractive index (Fig. \ref{fgr:tuning}a-b).
We track three higher order modes of the perturbed cavity shown in Figure \ref{fgr:tuning}a as Modes A, B and C, as they maintain their pattern, but linearly red-shift in resonant wavelength by 6 nm for $\Delta n = 0.005$ change in refractive index (Fig. \ref{fgr:tuning}d). After synthesis of the results of all simulations with GaAs thermal dependence of the refractive index \cite{talghader1995thermal}, the projected resonance tuning sensitivity is found to be on the order of nm/V$^2$.

For integration in complex circuits, we design bus waveguides for evanescent coupling to each cavity. Waveguide-integrated photonic crystal systems are simulated using Finite-Difference Time Domain simulations (Lumerical FDTD software) of the electromagnetic field propagating through the devices, with theoretical cross-sectional dimensions of 800 nm (width) by 150 to 220 nm (thickness). In Figure \ref{fgr:tuning}e we show a side profile of the 220 nm wide waveguide-cavity system showing the evanescent coupling to Mode B supported by the nanocavity. Figure \ref{fgr:tuning}f shows modeled transmission spectrum though the bus waveguide coupled to several higher order modes of the cavity. The transmission is calculated as the fraction of the modal power transmitted through the waveguide relative to the total generated source power. The waveguide thickness affects which modes the light will couple to more efficiently. The typical simulated transmission level was around -16 dB with 2 dB resonant contrast, while the typical intrinsic and loaded quality factors were $Q_0=2800$ and $Q_{loaded}=2000$, respectively.

\subsection*{Fabrication}

\begin{figure}[htbp]
\centering
\fbox{\includegraphics[width=\linewidth]{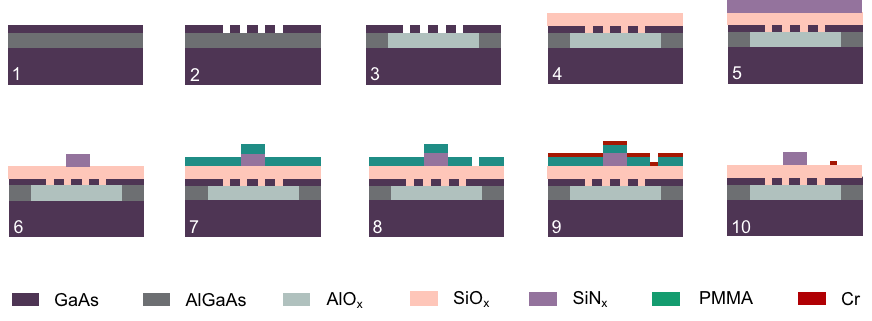}}
\caption{Electron-beam lithographic process for the implementation of the hybrid photonic architecture: suspended photonic crystal cavities are fabricated in GaAs on sacrificial AlGaAs substrate, further oxidized to AlO$_x$; HSQ is spun and thermally converted to SiO$_x$; SiN$_x$ is deposited in a PECVD process and fabricated into waveguides and grating couplers; Cr microheater are defined in a lift-off process using the PMMA e-beam resist.}
  \label{fgr:fab}
\end{figure}

Figure \ref{fgr:fab} illustrates the details of the three-step electron-beam lithographic process. In the first electron beam write, photonic crystals are implemented in a 130 nm GaAs slab on 800 nm aluminum gallium arsenide (AlGaAs) layer with 90\% aluminum content using standard 30 keV electron-beam lithography on ZEP 520A resist, followed by a reactive ion etch using boron trichloride (BCl$_3$), chlorine (Cl$_2$) and argon (Ar) plasma. The AlGaAs layer is subsequently converted to AlO$_x$ using an oxidation furnace at 423 $^{\circ}$C for 45 minutes, where the oxidation is facilitated by the air holes that are etched into the GaAs, and reach into the AlGaAs layer because of an intentional over-etch. The high aluminum content of the AlGaAs layer is specifically chosen to improve the efficiency of oxidation in an oxidation furnace. The reduced refractive index of the oxide layer allows light to be confined in the GaAs layer, and results in moderately high quality factor resonances ($Q \sim 2000$). After this oxidation step, a top glass cladding is formed for the devices using flowable oxide (Fox-15 Dow Corning) that fills in the air gaps at the photonic crystal lattice and forms a 200 nm thick SiO$_x$ layer on top of the devices \cite{debnath2013highly}. After curing the flowable oxide in a PECVD chamber at 350 $^{\circ}$C for one hour, a 220-nm layer of SiN$_x$ is deposited on the wafer in a PECVD chamber also at 350 $^{\circ}$C. As an alternative to this oxidation technique, we have also explored devices where the sacrificial AlGaAs layer is removed using hydrofluoric acid, resulting in a free-standing GaAs membrane, and the device is completely immersed in flowable oxide, followed by high temperature curing. Although typical devices have higher quality factors because of the symmetric oxide cladding, this technique is found to be less robust for applications that require scaling because of residual stress and bending in some devices. Next, using hydrogen silsesquioxane (HSQ) as an electron-beam resist on the oxide layer, waveguides are overlaid on top of the photonic crystal cavity in a second e-beam lithography step. These waveguides potentially allow us to address individual or multiple cavities simultaneously in a transmission measurement. \cite{debnath2013highly}. In the case of multiple cavity systems a set of holes is etched in the first lithography step bridging the regions between adjacent cavities. These bridges serve a dual role, preventing sudden changes in height (steps) for the bus waveguide at the edges of the photonic crystal regions resulting from the multi-step deposition process, and also providing entry ways for oxidizing the AlGaAs layer underneath the bridge region. This oxidation is critical for light to propagate with low losses between distant nodes.

The waveguides are etched using HBr plasma. Semicircular gratings are used as an input/output interface with free-space light. A 50 nm thick Cr microheater layer is lifted-off using a 100 nm thick Poly-methyl methacrylate (PMMA) A4 electron-beam resist in the third ebeam lithography step.

\begin{figure}[htbp]
\centering
\fbox{\includegraphics[width=\linewidth]{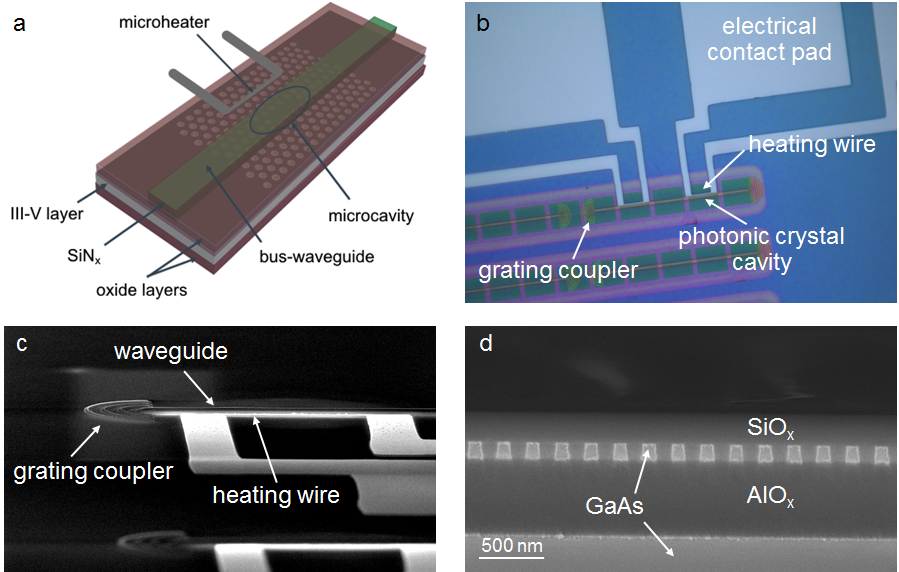}}
\caption{a) An illustration of the layers in the implemented hybrid photonic architecture. b) A microscope image of the integrated system. c) An SEM image of the integrated system. d) A cross-sectional SEM image showing photonic crystal mirrors buried between two oxide layers.}
\label{fgr:sems}
\end{figure}

A schematic view of a single integrated device is shown in Figure \ref{fgr:sems}a. However, the true potential of the platform is realized when multiple cavities are connected via a common bus waveguide. Images of fabricated multi-node systems are shown in Figure \ref{fgr:sems}b-d. The device in Figure \ref{fgr:sems}b shows devices where several photonic crystal cavities are addressed using a single bus waveguide, and two of these cavities can be individually addressed using local heating, for potentially controlling the relative phase between the two cavities through simultaneous frequency tuning. Because of the integrated waveguide that sits vertically on top of the cavity, the microheaters are laterally shifted from the cavity by at least half the width of the waveguide resulting in a poorer thermal tuning efficiency compared to heaters directly above the cavity; however this issue can be compensated by a voltage change in the electrical driving. The scanning electron micrograph (SEM) in Figure \ref{fgr:sems}c shows the surface of the device with the grating coupler, waveguide, and heating elements visible. The photonic crystal is embedded underneath the surface. The cross-sectional SEM image in Figure \ref{fgr:sems}d, obtained by cleaving through a device, shows the vertical layers, and the uniformity of the integration process. The figure shows that the PhC holes get completely filled with SiO$_x$ that has a smooth interface with AlO$_x$ layer.

\section*{Experimental results}

\begin{figure}[htbp]
\centering
\fbox{\includegraphics[width=8cm]{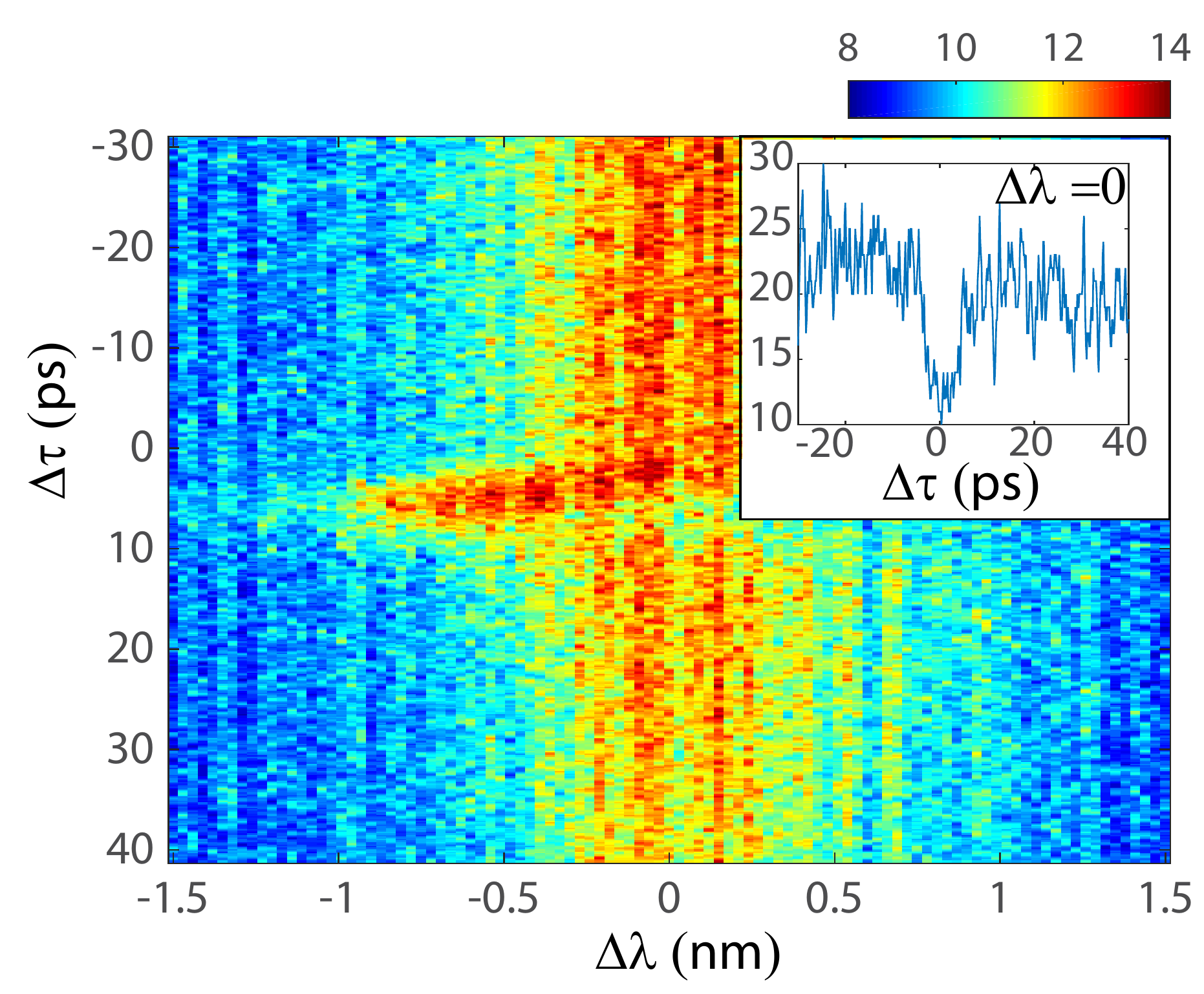}}
\caption{a) Free-carrier dynamics of oxide-clad photonic crystal device using a pump-probe measurement. The plot shows 10log$_{10}$ of the reflected probe laser intensity and the resonant-probe condition is plotted inset.}
\label{fgr:nonlinear}
\end{figure}

We characterize the developed hybrid photonics system in a confocal microscope under orthogonal incidence. An important result to verify is the nonlinear performance of our devices after the oxide cladding. For this experiment, sub-picosecond pulses from a Ti:sapphire laser at 820 nm are incident on the top of the oxide-clad cavity region and serve as a pump, while a continuous wave tunable laser (New Focus Velocity) is used to resonantly scan across the cavity mode wavelength at 901.5 nm, and serves as the probe. The reflected probe signal from the cavity is filtered using a spectrometer and measured using a streak camera (Hamamatsu) offering 2 ps dynamic resolution. In  Figure \ref{fgr:nonlinear}, we monitor the reflected probe signal from the cavity as a function of probe detuning from cavity resonance, expressed as $\Delta$$\lambda =$$\lambda$ -$\lambda_0$, and delay between pump and probe, expressed as $\Delta$$\tau$ for pump power of 200 $\mu$W. The arrival of the pump laser results in excitation of free carriers in the GaAs cavity region above the GaAs bandgap, leading to a change in refractive index through the free-carrier dispersion effect, and thus a blue-shift in the cavity resonance \cite{Bennett}. The switching time for the cavity is measured to be around 6 ps. These experiments confirm that the nonlinear effects that are observed in air-clad GaAs photonic crystal cavities are also evident in oxide-clad devices. Additional improvements in switching speeds could be pursued through atomic layer deposition of the cladding as a means of minimizing free carrier surface recombination \cite{moille2016integrated}.

\begin{figure}[htbp]
\centering
\fbox{\includegraphics[width=8cm]{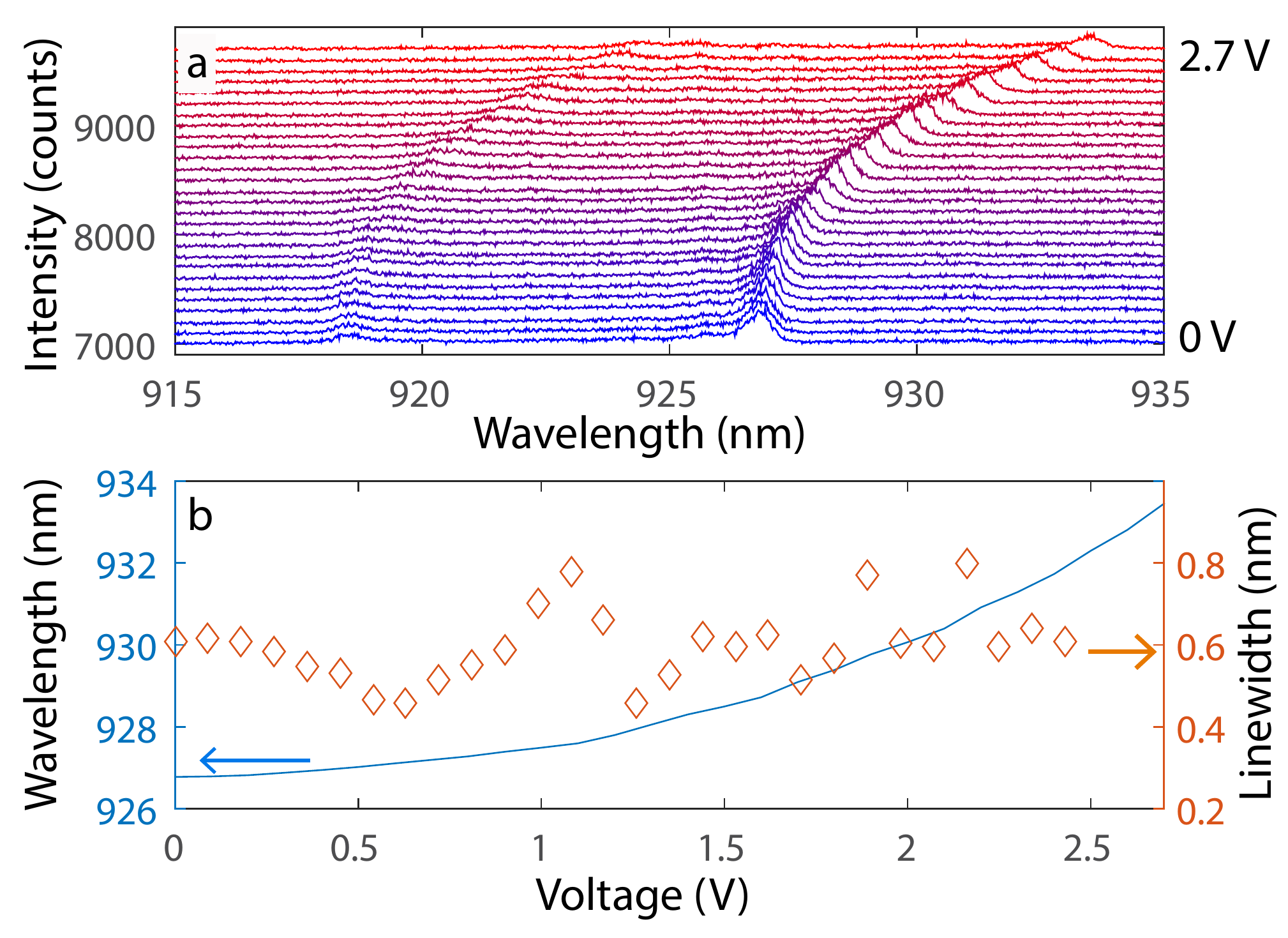}}
\caption{a) Thermal tuning of single photonic crystal cavity resonances across a 7 nm range. b) The wavelength and the linewidth of the tuned 926 nm resonance.}
\label{fgr:exp1}
\end{figure}

After the nonlinear measurements, we subsequently test the performance of the microheaters in tuning the resonance of the cavity mode. Electronic DC probes with a 60 $\mu$m pitch are contacted with the microheater pads, and the resulting I-V curve is analyzed to determine the microheater resistance, usually found to be 4 k$\Omega$. As a first step, we tune a single linear L3 defect cavity node using a broadband light source at normal incidence to track the cavity mode. The cavity spectrum exhibits two resonances at 919 nm (third order mode) and 926 nm (second order mode) with quality factor $Q \sim 1500$ at 0 V for the latter. As we increase the applied voltage, a clear redshift of the cavity spectra can be observed in experiments, consistent with thermal tuning. Figure \ref{fgr:exp1}a shows this spectral tuning under an applied voltage range of 0 to 2.7 V, where we acquire the cavity spectrum at regular intervals while tuning the voltage. We fit the acquired data using nonlinear curve fitting in Matlab and track the resonant wavelength and linewidth of the higher wavelength mode as we tune the voltage. The results are shown in  Figure \ref{fgr:exp1}b. We observe overall tuning of nearly 7 nm. The tuning is continuous and reversible, with the sensitivity of 1 nm/V$^2$, as modeled. In the data, the linewidth of the cavity mode seems to fluctuate over the tuning range, however these variations are most likely an artifact of the free-running acquisition scheme which continuously tunes the resonance throughout the signal collection affecting its lineshape. These fluctuations are therefore observed especially at higher voltages where the slope of the shift increases. Overall, the tuning does not degrade the quality factor significantly.

\begin{figure}[htbp]
\centering
\fbox{\includegraphics[width=8cm]{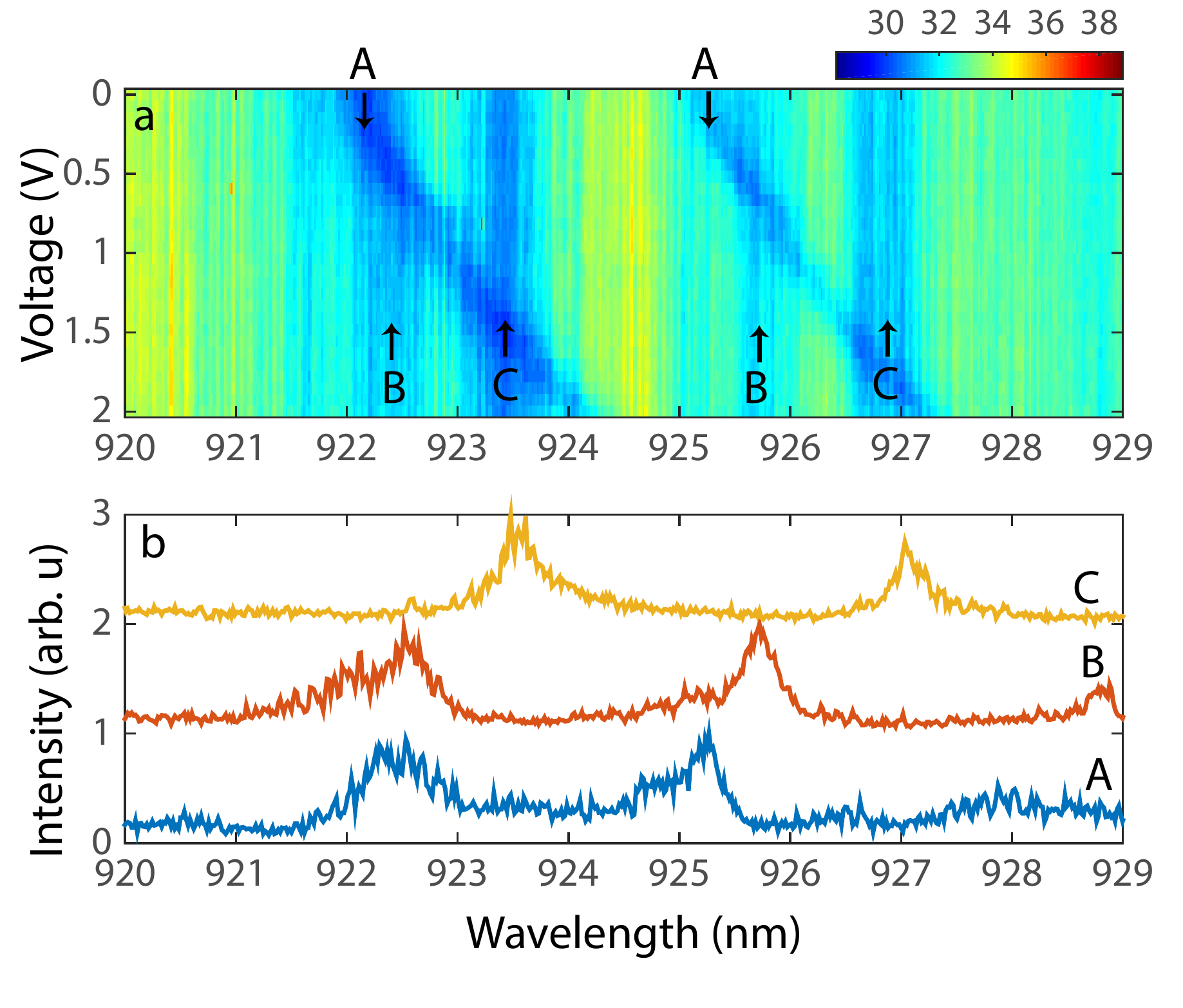}}
\caption{a) Transmission spectra for a device comprising of three cavities, one of which is continuously thermally tuned and shows a red shift with increasing voltage. The colors represent 10log$_{10}$(transmission) (dB). b) Scattering spectra for all three cavities at 0 V.}
\label{fgr:exp2}
\end{figure}

As a final demonstration of the scalable hybrid tunable architecture, we study three perturbed photonic crystal cavities that are separated by 11 $\mu$m each and connected using a common bus waveguide. Light is incident at an input grating coupler (10 dB loss) in the waveguide plane and transmitted through the bus waveguide, evanescently coupling to each resonance. The transmission contrast is limited to < 3 dB by an imperfect mode and phase matching between the waveguide and cavity modes  \cite{melchiorri2005propagation}. The total transmission loss is estimated at above 30 dB, and can be significantly improved with the use of more efficient grating couplers \cite{sapra2018inverse}. Figure \ref{fgr:exp2}a shows tuning spectra where one of the three cavities (whose two modes red-shift in the plot) is tuned through the resonances of several modes from other cavities that maintain their original wavelengths (as evidenced by their vertical traces). The lower three plots in figure \ref{fgr:exp2}b show spectra of the individual cavities obtained by focusing light at the input grating and observing the vertical scattering from the cavity at 0 V, with the lowest curve (blue) representing the cavity that is thermally tuned. These cavities exhibit resonances at A:(922.4, 925.3), B:(922.6,925.7), and C:(923.5, 927.1) nm, where A is the cavity being tuned, while cavities B and C do not experience any resonant shift. In this case, the cavity modes appear as peaks, in contrast to the transmission measurement where a dip is observed for each resonance. The resonances of these identically designed cavities are initially detuned because of nanofabrication errors. As cavity A is tuned using the microheater, we see mode crossings at several voltages. The lower wavelength mode of cavity A is resonant with a mode of cavity B at 0.2 V, and with cavity C at 1.6 V. Similarly the higher wavelength mode of cavity A is resonant with modes of cavity B and C at 0.7 and 1.7 V, respectively. The transmission plot shows that the tuning range of the microheaters can overcome fabrication discrepancies, and also confirms that the tuning scheme is local to the immediate region around a single cavity. With the demonstrated tuning voltages and the device periodicities, the thermal dissipation flux corresponds to 1 kW/cm$^2$, which is routinely achieved using liquid cooling techniques \cite{mudawar2000assessment}.

In summary, we have implemented a thermally tunable hybrid photonic architecture suitable for coherent interaction of free-space light and tunable GaAs photonic crystal cavities facilitated by SiN$_x$ bus-waveguide terminated by grating couplers. We show excellent thermal tuning performance using custom-designed microheaters. Importantly, we show that the tuning scheme, in addition to being reversible, is also local to the region near the photonic crystal, and does not result in degradation of the cavity mode or waveguide. This experiment demonstrates tuning control for aligning arbitrarily distant optical cavities to the same optical resonance. Simultaneous wavelength tuning at a second node, or the addition of a heater adjacent to the waveguide region will allow us to also control the phase of light in the optical circuit. This experiment therefore provides a framework for designing  highly nonlinear, coherent optical circuits \cite{santori2014quantum, 7738704}.

\begin{acknowledgement}

This work is supported by the Defense Advanced Research Projects Agency under Agreement No. N66001-12-2-4007. The authors acknowledge contributions from Jason Pelc and Charles M. Santori, and useful consultations with Professor Jelena Vu\v ckovi\' c.

\end{acknowledgement}

\bibliography{heaterspaper}

\end{document}